\documentstyle[amssymb,preprint,prb,aps]{revtex}
\begin{document}

\begin{titlepage}

\title
{Energy levels and far-infrared spectroscopy for two electrons in a semiconductor 
nanoring}

\author{Hui Hu\footnote {Email: huhui@Phys.Tsinghua.edu.cn} }
\address{Department of Physics, Tsinghua University, Beijing 100084, P. R. China}
\author{Jia-Lin Zhu} 
\address{Department of Physics, Tsinghua University, Beijing 100084, P. R. China \\
Center for Advanced Study, Tsinghua University, Beijing 100084, P. R. China}
\author{Jia-Jiong Xiong}
\address{Department of Physics, 
Tsinghua University, Beijing 100084, P. R. China}

\maketitle
\begin{abstract}
The effects of electron-electron interaction of a two-electron nanoring on
the energy levels and far-infrared spectroscopy have been investigated
based on a model calculation which is performed within the exactly numerical
diagonalization. It is found that the interaction changes the energy spectra 
dramatically, and also shows significant influence on the far-infrared spectroscopy. 
The crossings between the lowest spin-singlet and triplet states induced by 
the coulomb interaction are clearly revealed. Our results are related to the 
experiment recently carried out by A. Lorke {\em et al.} [Phys. Rev. Lett. 84, 2223 (2000)].

\noindent
{\bf PACS number(s)}:  71.15.-m, 73.61.-r, 78.30.-j, 73.20.Dx
\end{abstract}

\end{titlepage}

\section{Introduction}

Rapid progress in nanostructure technology has made it possible to fabricate
various nanometer quantum devices which have potential applications in
microelectronics. Such ultrasmall devices contain only a few electrons and
the electron-electron interaction is proposed to be of great importance to
theirs energy level structures and optical properties. It leads to a number
of new quantum phenomena. One of the most interesting phenomena is the {\em %
spin oscillation} of the ground state in an external magnetic field, which
is due to the interplay between three energies: the confinement potential,
the Zeeman energy, and the electron-electron interaction. The simplest case,
a pillar few electrons quantum dot (QD) with a parabolic potential, has been
extensively investigated, and the spin oscillation in a QD is well
understood. \cite{wagner,zhu97a,dineykhan} The theoretical predictions are
also confirmed in the experiment through the conductance measurements in the
finite drain-source voltage regime. \cite{kouwen}

The semiconductor quantum ring is another interesting example. In 1993, D.
Mailly {\em et al}. measured the {\em persistent currents }in a mesoscopic
single GaAs ring induced by a magnetic flux threading the interior of ring, 
\cite{mailly} and the experimental results have attracted a lot of
theoretical interests. One of the basic questions addressed by many
theoretical explanation is concerned with the role of the electron-electron
interaction. \cite{chen,yi,ben,stafford,wend95a,wend95b,wend95c,zhu97b} For
a {\em narrow-width} rings, an adiabatic approximation allows one to
decouple the radial motion from the angular motions and to arrive at
analytical solutions for the wave functions and the energy spectra. \cite
{wend95c,wend96} As shown by L. Wendler {\em et al.}, \cite{wend95c,wend96}
the interplay of the Coulomb repulsion between the electrons and the
confining potential forms a relatively rigid rotator with internal azimuthal
excitations and confined radial motions, {\em i.e.}, the picture of a
rotating Wigner molecule. However, at the moment, the influence of the
electron-electron interaction in a {\em finite-width} and {\em nanoscopic}
quantum ring, {\em i.e.}, nanoring, is still less well understood. \cite
{chak,gudm,halonen}

Very recently, using the self-assembly techniques, A. Lorke and
collaborators demonstrate the realization of nanoscopic semiconductor
quantum rings inside a completed field-effect transistor (FET) structure. 
\cite{lorke,emperador,warburton} Quite different from the conventional
sub-micron mesoscopic structures, the nanorings are in the {\em true }%
quantum limit. By applying two complementary spectroscopic techniques,
capacitance-voltage (CV) spectra and far-infrared (FIR) spectroscopy, they
investigate both the ground state transition and excitation's properties of
these two-electron nanorings in a magnetic field perpendicular to the plane
of rings. Although the main experimental results can be qualitatively
explained by the single electron picture, some contradictions remain, i.e.
the coulomb interaction energy estimated roughly $20$ {\rm meV}, is too
large to be ignored safely. Hence, more in-depth theoretical works are
desirable, especially in view of the existences of the very strong
electron-electron coulomb interaction.

In this paper, we would like to study the energy levels and FIR spectroscopy
of a two-electron nanoring, and pay special attention to the effects of the
Coulomb interaction. First of all, for a ring-like confinement potential,
the total Hamiltonian cannot be separated into the center-of-mass and the
relative-motion terms. We develope a new theoretical method to handle this
non-separability. It consists of the well-known series solution method, \cite
{zhu97a,zhu98} which is effective to solve the single particle problem, and
the exact diagonalization method. \cite{eto,ugajin} On the other hand, we
show that the electron-electron interaction can change the energy levels
significantly. An obvious feature induced by the interaction is the
intersection between the lower levels. It presents the spin oscillation of
the ground state in an external field. Moreover, the results obtained by A.
Lorke {\em et al.} can be explained more realistically in our model.
Further, the generalized Kohn theorem breaks down in the FIR spectroscopy
due to the mixing between the center-of-mass and the relative-motion modes
as mentioned above. \cite{oh,magnus} We outline here two prominent features
of the FIR spectroscopy caused by electron-electron interaction: splitting
and dis-continuous drops of the resonance energies, which are suggested to
be detectable with circularly polarized far-infrared light.

The organization of this paper is as follows.\ In the next section, we
introduce the model and Hamiltonian briefly. In Sec III, the main procedures
of the numerical calculation are outlined. Sec IV is devoted to the results
and discussions, followed by a summary in Sec V.

\section{The model and Hamiltonian}

The nanoring is considered to have two electrons with an effective
conduction-band-edge mass $m_e^{*}$ moving in the $x-y$ plane, and a
ring-like confining potential can be introduced $U(\vec{r})=\frac 12%
m_e^{*}\omega _0^2\left( r-R_0\right) ^2$, \cite{halonen,lorke,emperador}
where $\omega _0$ is the characteristic frequency of the radial confinement
and $R_0$ is the ring's radius. This system is subjected to a perpendicular
uniform magnetic field which is described by a vector potential $\vec{A}%
\left( \vec{r}\right) =\frac 12\vec{B}\times \vec{r}$ in the symmetric (or
circular) gauge. The resulting Hamiltonian is given by 
\begin{equation}
{\cal H}=\sum\limits_{i=1,2}\left\{ \frac 1{2m_e^{*}}\left( \vec{p}_i+e\vec{A%
}\left( \vec{r}_i\right) \right) ^2+U(\vec{r}_i)\right\} +\frac{e^2}{4\pi
\varepsilon _0\varepsilon _r\left| \vec{r}_1-\vec{r}_2\right| },
\label{Hami1}
\end{equation}
where $\vec{r}_i=\left( x_i,y_i\right) $ and $\vec{p}_i=-i\hbar \vec{%
\triangledown}_i$ are respectively the position vector and momentum operator
of the $i$-th electron with charge $-e$. $\varepsilon _0$ is the vacuum
permittivity and $\varepsilon _r$ is the static dielectric constant of the
host semiconductor. In addition, there also exist the spin interaction term
with the magnetic field ${\cal H}_{spin}=g^{*}\mu _B\left( \vec{S}_1+\vec{S}%
_2\right) \cdot \vec{B},$ where $g^{*}$ is the Land\'{e} factor, and $\mu _B$
is the Bohr magneton. Due to the small $g^{*}$ in semiconductor, in general, 
${\cal H}_{spin}$ is very small and can be ingored safely ({\em i.e.}, for $%
g^{*}=-0.44$ in GaAs materials and $B=10$ {\rm T}, the typical value of $%
{\cal H}_{spin}$ is $0.25$ {\rm meV}). It is worthy to note that in the
limit case $R_0=0$, the nanoring simply reduces to a parabolic quantum dot,
which can be solved trivially by separating the Hamiltonian into
center-of-mass and relative-motion terms. The appearance of $R_0$ breaks
this separability, and it becomes more complex.

We apply the exact diagonalization method by constructing the basis with the
single particle wavefunctions of Hamiltonian 
\begin{equation}
{\cal H}_s=\frac 1{2m_e^{*}}\left( \vec{p}+e\vec{A}\left( \vec{r}\right)
\right) ^2+U(\vec{r}).  \label{Hami2}
\end{equation}
These wavefunctions labelled by the radial quantum number $n$ and orbital
angular-momentum quantum number $m$ have the form 
\begin{equation}
\psi _{nm}\left( \vec{r}\right) =R_{nm}(r)\exp \left( im\varphi \right)
\qquad n=0,1,2...,\quad m=0,\pm 1,...,
\end{equation}
where the radial part $R_{nm}$ will be solved exactly by using the series
expansion method. \cite{zhu97a}

\section{Formula and calculation methods}

\subsection{Series solution}

For the sake of convenience, we use the effective atomic units, in which the
effective Rydberg $R_y^{*}=\frac{m_e^{*}e^4}{2\hbar ^2\left( 4\pi
\varepsilon _0\varepsilon _r\right) ^2}$ and the effective Bohr radius $%
a^{*}=\frac{4\pi \varepsilon _0\varepsilon _r\hbar ^2}{m_e^{*}e^2}$ are
taken to be the energy and length units, respectively. Then, the Hamiltonian
(\ref{Hami2}) has the form 
\begin{equation}
{\cal H}_s=-\vec{\triangledown}^2+\frac 14\gamma _b^2r^2+\frac 14\gamma
_d^2\left( r-R_0\right) ^2+\gamma _b\hat{L}_z,
\end{equation}
where the magnetic field $\gamma _b$ is measured in the unit $\frac{\hbar
\omega _c}{2R_y^{*}}$ with cyclotron frequency $\omega _c=\frac{eB}{m_e^{*}}$%
, $\gamma _b\hat{L}_z$ is the Zeeman term, and $\gamma _d^{-1/2}=\left( 
\frac{R_y^{*}}{\hbar \omega _0}\right) ^{1/2}$ is related to confinement
region of the electrons. It is interesting to note how large the units of
semiconductor materials are. For GaAs materials, for example, $R_y^{*}=5.8$ 
{\rm meV}$,$ $a^{*}=10$ {\rm nm}, and $\gamma _b=1$ corresponds to $B=6.75$ 
{\rm T}.

Now, we have to solve the Schr\H{o}dinger-like equation 
\begin{equation}
{\cal H}_s\left[ R_{nm}(r)\exp \left( im\varphi \right) \right]
=E_{nm}\left[ R_{nm}(r)\exp \left( im\varphi \right) \right]
\end{equation}
to obtain the energy $E_{nm}$ and radial part of wavefunction $R_{nm}(r)$.
It is easy to find the equation satisfied by the function $R_{nm}\left(
r\right) $%
\begin{equation}
\left\{ \frac{d^2}{dr^2}+\frac 1r\frac d{dr}+\left[ \left( E_{nm}-m\gamma
_b\right) -\frac{m^2}{r^2}-\frac 14\gamma _b^2r^2-\frac 14\gamma _d^2\left(
r-R_0\right) ^2\right] \right\} R_{nm}\left( r\right) =0.  \label{Rad}
\end{equation}
We are prevented from analytically exact solutions of the eigenvalue problem
because Eq. (\ref{Rad}) with suitable boundary conditions is beyond the
analytical problem of confluent hypergeometric equations. However, we can
use the method of series expansion to obtain the series forms in different
regions of Eq. (\ref{Rad}) and the exact values of $E_{nm}$. \cite
{zhu97a,zhu98}

It should be noted that $r=0$ and $r=+\infty $ are respectively the regular
and irregular singularity points of Eq. (\ref{Rad}). So it is natural to
divide the whole region $[0,+\infty )$ into three parts, $[0,+\infty
)=[0,r_0)\cup [r_0,r_\infty )\cup [r_\infty ,+\infty )$, where $r_0$ and $%
r_\infty $ are the two dividing points. With suitable adjustment of $r_0$
and $r_\infty $, very high numerical precision can be archived. In all the
three kinds of regions, the function $R_{nm}\left( r\right) $ is found to be
the form: 
\begin{equation}
R_{nm}\left( r\right) =\left\{ 
\begin{array}{cc}
Ar^l\sum\limits_{n=0}^\infty a_nr^n & \qquad 0\leqslant r<r_0, \\ 
C_i\sum\limits_{n=0}^\infty c_{in}\left( r-R_i\right)
^n+D_i\sum\limits_{n=0}^\infty d_{in}\left( r-R_i\right) ^n & \qquad
R_i\leqslant r<R_{i+1}, \\ 
B\exp \left( -\frac 14\gamma r^2+\frac{\gamma _d^2}{2\gamma }R_0r\right)
r^s\sum\limits_{n=0}^\infty b_nr^{-n} & \qquad r_\infty \leqslant r<+\infty ,
\end{array}
\right.
\end{equation}
where $\gamma =\sqrt{\gamma _b^2+\gamma _d^2}$, $l=\left| m\right| $ and $s=%
\frac{\left( E_{nm}-m\gamma _b\right) }\gamma -1-\frac{\gamma _b^2\gamma _d^2%
}{4\gamma ^3}R_0^2$. In order to improve precision, we further divide the
region $[r_0,r_\infty )$ into $N$ pieces, denoted by $R_i$ ($i=1,...,N$),
here $R_1=r_0$ and $R_N=r_\infty $. $A$, $C_i$, $D_i$ and $B$ are constants, 
$a_n$, $c_{in}$, $d_{in}$ and $b_n$ are expanding coefficients and can be
determined by the recurrence relation coming from Eq. (\ref{Rad}), the
initial values of coefficients are chosen to be $a_0=1$, $c_{i0}=d_{i1}=1$, $%
c_{i1}=d_{i0}=0$ and $b_0=1$.

Using the matching conditions at $r=R_i$ ($i=1,...,N$), and the $2\times 2$
transfer matrices, we can deduce the equation for eigenenergies $E_{nm}$
easily. Then the constant $A$, $C_i$, $D_i$ and $B$ can be evaluated by
normalization condition, and $R_{nm}\left( r\right) $ is obtained
numerically.

To close this subsection, it is interesting to point out that the method
mentioned above is very suitable for numerical calculations, and can be
modified to handle various differential equations similar to Eq. (\ref{Rad}).

\subsection{Exact diagonalization}

Once the single particle wavefunctions are obtained, we go ahead to
construct the basis of the two-electron wavefunctions. It is obvious that
the total $z$ component of the angular momentum operator of two electrons, $%
\hat{L}_{z,total}=\hat{L}_{z1}+\hat{L}_{z2}$, is a constant of motion, i.e., 
$\left[ {\cal H},\hat{L}_{z,total}\right] =0$ is valid, from which the
rotational invariance of the problem follows. Thus, one can work in one
subspace labelled by a good quantum number $L$ instead of the whole huge
Hilbert space. We denote the corresponding two-electron wavefunction by $%
\psi _L\left( \vec{r}_1\sigma _1;\vec{r}_2\sigma _2\right) $ which depends
on the spatial coordinates $\left\{ \vec{r}_i\right\} $ and the spin
coordinates $\left\{ \sigma _i\right\} $ ($i=1,2$). Because the Hamiltonian
of Eq. (\ref{Hami1}) does not depend on the spin operator, in our two
electrons case, the wavefunction can separate into the orbital part $\psi
_L\left( \vec{r}_1,\vec{r}_2\right) $ and the spin part $\chi _S\left(
\sigma _1,\sigma _2\right) $: 
\begin{equation}
\psi _{LS}\left( \vec{r}_1\sigma _1;\vec{r}_2\sigma _2\right) =\psi
_L^{\lambda _1\lambda _2}\left( \vec{r}_1,\vec{r}_2\right) \chi _S\left(
\sigma _1,\sigma _2\right)  \label{wvf}
\end{equation}
with 
\begin{equation}
\psi _L^{\lambda _1\lambda _2}\left( \vec{r}_1,\vec{r}_2\right) =c\left[
\psi _{n_1m_1}\left( \vec{r}_1\right) \psi _{n_2m_2}\left( \vec{r}_2\right)
+\left( -\right) ^S\psi _{n_2m_2}\left( \vec{r}_1\right) \psi
_{n_1m_1}\left( \vec{r}_2\right) \right] ,
\end{equation}
where the normalized wavefunction $\psi _{LS}\left( \vec{r}_1\sigma _1;\vec{r%
}_2\sigma _2\right) $ is further labelled by the total spin number $S=0$ or $%
1$, corresponding to the singlet and triplet state. $\lambda _i$ $\left(
i=1,2\right) $ stands for the quantum number pair $\left( n_im_i\right) .$ $%
L=m_1+m_2$. $\psi _{nm}\left( \vec{r}\right) $ is the single particle
wavefunction. $c=\sqrt{\frac 12\text{ }}$ as $\left( n_1m_1\right) \neq
\left( n_2m_2\right) ,$ and $c=\frac 12$ as $\left( n_1m_1\right) =\left(
n_2m_2\right) .$ Obviously, the wavefunction constructed above satisfies the
antisymmetric condition, 
\begin{equation}
{\cal P}_{12}\psi _{LS}\left( \vec{r}_1\sigma _1;\vec{r}_2\sigma _2\right)
=-\psi _{LS}\left( \vec{r}_1\sigma _1;\vec{r}_2\sigma _2\right) ,
\end{equation}
where ${\cal P}_{12}$ is the permutation operator.

In the next step, we diagonalize the Hamiltonian (\ref{Hami1}) numerically
in a restricted configuration space. The space is constructed by choosing
the wavefunctions having the form (\ref{wvf}) in the lowest $f$ levels. The
secular equation of finite degree $f$ is given by 
\begin{equation}
\det \left\| \left( E_i^{(0)}-E\right) \delta _{ij}+\Delta _{ij}\right\|
=0,\qquad i,j=1,...,f,  \label{se}
\end{equation}
where $E_i^{(0)}=E_{n_1^im_1^i}+E_{n_2^im_2^i}$ is the single particle
energy, $i=\left( n_1^im_1^i;n_2^im_2^i\right) $ and $j=\left(
n_1^jm_1^j;n_2^jm_2^j\right) $ represent the quantum number levels. $\Delta
_{ij}$ is the matrix element of electron-electron interaction in unit of $%
R_y^{*}$%
\begin{equation}
\Delta _{ij}=\left\langle i\left| \frac 2{\left| \vec{r}_1-\vec{r}_2\right| }%
\right| j\right\rangle =\Delta _{ij}^c+\left( -\right) ^S\Delta _{ij}^e,
\end{equation}
where 
\begin{equation}
\Delta _{ij}^c=\int \int d\vec{r}_1d\vec{r}_2\psi _{n_1^im_1^i}^{*}\left( 
\vec{r}_1\right) \psi _{n_2^im_2^i}^{*}\left( \vec{r}_2\right) \frac 2{%
\left| \vec{r}_1-\vec{r}_2\right| }\psi _{n_1^jm_1^j}\left( \vec{r}_1\right)
\psi _{n_2^jm_2^j}\left( \vec{r}_2\right) ,
\end{equation}
and 
\begin{equation}
\Delta _{ij}^e=\int \int d\vec{r}_1d\vec{r}_2\psi _{n_1^im_1^i}^{*}\left( 
\vec{r}_1\right) \psi _{n_2^im_2^i}^{*}\left( \vec{r}_2\right) \frac 2{%
\left| \vec{r}_1-\vec{r}_2\right| }\psi _{n_2^jm_2^j}\left( \vec{r}_1\right)
\psi _{n_1^jm_1^j}\left( \vec{r}_2\right) .
\end{equation}
$\Delta _{ij}^c$, $\Delta _{ij}^e$ can be computed numerically from the
series solution of $\psi _{nm}\left( \vec{r}\right) $.

By diagonalizing the secular equation (\ref{se}) in each subspace $(S,L)$,
we obtain the $m$-th energy level $E_m$ and the corresponding two-electron
wave-function: 
\begin{equation}
\Phi _m\left( \vec{r}_1\sigma _1;\vec{r}_2\sigma _2\right)
=\sum\limits_{\lambda _1\lambda _2}A_{\lambda _1\lambda _2}^m\psi
_L^{\lambda _1\lambda _2}\left( \vec{r}_1,\vec{r}_2\right) \chi _S\left(
\sigma _1,\sigma _2\right) .  \label{wv-function}
\end{equation}

\subsection{Optical absorption}

In the electronic dipole approximation, \cite{wend96} the absorption
coefficient is given by 
\begin{equation}
\alpha \left( \omega \right) =c\omega \sum\limits_{fi}\left| \left\langle
f\left| \vec{e}\cdot \vec{d}\right| i\right\rangle \right| ^2\delta \left(
\omega -\omega _{fi}\right) \left( f_i^{(0)}-f_f^{(0)}\right) ,
\label{alpha1}
\end{equation}
where $\vec{e}$ is the complex polarization vector of the spatially constant
external electronic field, $\vec{d}=-\left( \vec{r}_1+\vec{r}_2\right) $ is
the electronic dipole operator of two electrons, and $f^{(0)}$ is the
equilibrium Fermi distribution. The summation is over all the two-electron
eigenstates, $\omega _{fi}$ is proportional to the energy difference between
the initial state $\mid i\rangle $ and final state $\mid f\rangle :$ $\omega
_{fi}=\omega _f-\omega _i=\left( \frac{E_f-E_i}\hbar \right) .$ $c$ is a
constant factor. Restricting ourselves to zero temperature $T=0$K, Eq. (\ref
{alpha1}) reduces to 
\begin{equation}
\alpha \left( \omega \right) =c\omega \sum\limits_f\left| \left\langle
f\left| \vec{e}\cdot \vec{d}\right| 0\right\rangle \right| ^2\delta \left(
\omega -\omega _{f0}\right) ,  \label{alpha2}
\end{equation}
where $\mid 0\rangle $ and $\mid f\rangle $ represent the ground state and
excited state respectively.

For circularly polarized light we have $\vec{e}=\sqrt{\frac 12}\left( 1,\pm
i\right) ,$ from which it follows $\vec{e}\cdot \vec{d}=-\sqrt{\frac 12}%
\left( r_1e^{\pm i\varphi _1}+\left( 1\rightarrow 2\right) \right) $. Using
Eq. (\ref{wv-function}), we obtain 
\begin{equation}
\left\langle f\left| \vec{e}\cdot \vec{d}\right| 0\right\rangle =-\sqrt{%
\frac 12}\sum\limits_{\lambda _1^0\lambda _2^0}\sum\limits_{\lambda
_1^f\lambda _2^f}A_{\lambda _1^f\lambda _2^f}^{f*}A_{\lambda _1^0\lambda
_2^0}^0\left( \left( d_{\lambda _1^0\lambda _1^f}^{\pm }\delta _{\lambda
_2^0\lambda _2^f}+\left( -\right) ^Sd_{\lambda _1^0\lambda _2^f}^{\pm
}\delta _{\lambda _2^0\lambda _1^f}\right) +\left( 1\rightarrow 2\right)
\right) ,  \label{dipole}
\end{equation}
where the single particle matrix elements $d_{\lambda \lambda ^{\prime
}}^{\pm }$ are defined by 
\begin{equation}
d_{\lambda \lambda ^{\prime }}^{\pm }=\left\langle \lambda ^{\prime }\left|
r\exp \left( \pm i\varphi \right) \right| \lambda \right\rangle =\delta
_{m\pm 1,m^{\prime }}\int\limits_0^\infty r^2R_{\lambda ^{\prime }}\left(
r\right) R_\lambda \left( r\right) dr
\end{equation}
and $\left| \lambda \right\rangle $ represents the single particle
wave-function $\psi _{nm}\left( \vec{r}\right) $. It is obvious that the
absorption coefficient satisfies the dipole selection rule $\Delta L=\pm 1.$
Substituting Eq. (\ref{dipole}) into Eq. (\ref{alpha2}) and taking 
\begin{equation}
\delta \left( \omega -\omega _{f0}\right) =\frac{\Gamma /\pi }{\left( \omega
-\omega _{f0}\right) ^2+\Gamma ^2},
\end{equation}
where $\Gamma $ is a phenomenological broadening parameter, we arrive at 
\begin{eqnarray}
\alpha ^{\pm }\left( \omega \right) &=&c\frac \omega 2\sum\limits_f\left|
\sum\limits_{\lambda _1^0\lambda _2^0}\sum\limits_{\lambda _1^f\lambda
_2^f}A_{\lambda _1^f\lambda _2^f}^{f*}A_{\lambda _1^0\lambda _2^0}^0\left(
\left( d_{\lambda _1^0\lambda _1^f}^{\pm }\delta _{\lambda _2^0\lambda
_2^f}+\left( -\right) ^Sd_{\lambda _1^0\lambda _2^f}^{\pm }\delta _{\lambda
_2^0\lambda _1^f}\right) +\left( 1\rightarrow 2\right) \right) \right|
^2\times  \nonumber \\
&&\frac{\Gamma /\pi }{\left( \omega -\omega _{f0}\right) ^2+\Gamma ^2},
\end{eqnarray}
where $\pm $ corresponds to right and left circularly polarized lights,
respectively.

To check the numerical accuracy of the calculations we have used the $f$-sum
rules for the dipole operators, which can be expressd in terms of
ground-state quantities: \cite{pines,serra} 
\begin{equation}
\int\limits_0^\infty \left( \alpha ^{+}\left( \omega \right) +\alpha
^{-}\left( \omega \right) \right) d\omega =\left\langle 0\right| \left[
\left( \vec{e}\cdot \vec{d}\right) ^{+},\left[ {\cal H},\left( \vec{e}\cdot 
\vec{d}\right) \right] \right] \left| 0\right\rangle =N  \label{f-sum}
\end{equation}
in the effective atomic units, where $c$ is taken to be 1, and $N=2$ is the
number of electrons.

It is important to point out that in our case, due to non-separability of
the center-of-mass and the relative-motion modes, the generalized Kohn
theorem, which means that FIR can only be used to excite the center-of-mass
modes of electrons parabolically confined in circular quantum dot, will not
be held further. Thus we expect our FIR absorption result may reflect an
excitation of the relative motion of two electrons.

\section{Results and discussions}

To explain the experimental measurements, \cite{lorke} we have taken the
material parameters $\varepsilon _r=12.4$ and $m_e^{*}=0.067m_e$ for GaAs in
our calculations. The radial confinement strength $\hbar \omega _0$ and the
ring's radius $R_0$ are chosen to be $12$ {\rm meV} and $14$ {\rm nm}
respectively. The corresponding width of ring is about $15$ {\rm nm}, which
means that the electrons are confined in a wide ring. Thus in contrast to
the rotating Wigner molecule picture in a narrow-width quantum ring, \cite
{wend95c,wend96} the more pronounced energy spectra and optical properites
are expected. For the calculation in each $(S,L)$ subspace, we first solve
the single particle problem and save several hundreds single particle
states, then pick up the suitable single particle states to construct
thousands of two-electron states, among which only the lowest $f$ energy
levels are selected. Here we simply point out that our numerical
diagonalization scheme is very efficient and essentially exact in the sense
that the accuracy can be improved as desired by increasing $f$. For
instance, for the ground state in $\left( 0,0\right) $ subspace, the use of
64 basis states allows the precision to be within the relative convergence
of $\sim 10^{-4}.$ On the other hand, by checking the $f$-sum rule, we find
that the relative error of our FIR calculation is less than {\rm 1\%}.

\subsection{Spin oscillation}

The energy levels of two electrons in a nanoring as a function of the
magnetic field have been plotted in Fig.1. As mentioned above, only the
total spin and total angular momentum are conserved in our model, and then
for the sake of clearness, we only plot the lowest level in each $\left(
S,L\right) $ subspace$.$ In order to understand the role of the
electron-electron interaction in the two-electron spectra better, we first
describe the characteristics of the energy levels of two electrons in the
nanoring without interaction. As shown in Fig. 1(a), for small magnetic
field, the spectra have the characteristics of disk-like quantum dot, since
the electrons are confined in a wide nanoring whose radius is comparable to
its effective width. As the magnetic field increases from zero, there are
minima for the states with negative angular momentum $L$. These minima are
caused by the interplay between the Zeeman term and the ring-like
confinement potential. Moreover, the level with $L=0$ increases
monotonically, and changes more dramatically than the others. This leads to
a ground state transition from $L=0$ to $L=-2$ around $B=8$ {\rm T}, which
also reflects the fact that the nanoring becomes more and more narrow with
increasing the magnetic field. In the sufficient high field regime, it is
quite safe to speculate that the levels will be the same as the levels in
one-dimensional ring under a uniform magnetic field.

The electron-electron interaction can significantly change the
characteristics of the spectra described above. As shown in Fig. 1(b),
clearly visible is the ground state transitions around $B=3$, $7$ and $10$ 
{\rm T}. These transitions are quite different from that mentioned in the
previous paragraph:\ not only the total angular momentum but also the total
spin are changed. They present the spin-singlet-spin-triplet oscillation of
the ground state in the magnetic field, i.e., $\left( 0,0\right) \rightarrow
\left( 1,-1\right) \rightarrow \left( 0,-2\right) \rightarrow \left(
1,-3\right) $ states and so on. This phenomenon is indeed qualitatively
similar to that seen in a quantum dot. \cite{zhu97a}

For a better understanding of the singlet-triplet oscillation, it is
interesting to study the electron-electron interaction energies $E_r$,
defined by the difference between the energy $E$ with interaction and $%
E^{(0)}$ without interaction. In Fig. 2, the $E_r$ for different states are
plotted as a function of the magnetic field. It is readily seen that the $%
E_r $ increase with the magnetic field and the ordering is as follows: $%
E_r\left( 0,-1\right) >E_r\left( 0,0\right) >E_r\left( 0,-3\right)
>E_r\left( 1,-2\right) >E_r\left( 1,0\right) >E_r\left( 1,-1\right)
>E_r\left( 0,-2\right) >E_r\left( 1,-3\right) ...$. Compared with the
corresponding results in quantum dot, \cite{zhu97a} we find that the
ordering depends significantly on the form of the confinement potential,
e.g., the ordering $E_r\left( 0,0\right) >E_r\left( 0,-1\right) $ in quantum
dot is reversed in our case. However, the ordering $E_r\left( 0,0\right)
>E_r\left( 1,-1\right) >E_r\left( 0,-2\right) >E_r\left( 1,-3\right) $ is
still preserved, thus the trivial crossover around $B=8$ {\rm T} in Fig.
1(a) moves to the position $B=3$ {\rm T} in Fig. 1(b).

We comment on some other effects caused by the interaction. ($i$) When the
electron-electron interaction is excluded, the states $\left( 0,L\right) $
and $\left( 1,L\right) $ $\left( L=-1,-3\right) $ are degenerate in the
whole regime of the magnetic field. This degeneracy can be understood by the
fact that those states are constructed with the same single particle states,
the only difference between them is spin part which have no influence to
energy levels in the absence of interaction. When the electron-electron
interaction turns on, the whole energy levels are shifted to high energies
due to the repulsive coulomb interaction and the degenerated levels split. ($%
ii$) In the high magnetic field regime, the energy levels appear to be more
separated in the presence of interaction, and it indicates that the
interaction effect becomes larger in this regime.

\subsection{FIR spectroscopy}

In the previous subsection, we have demonstrated that the electron-electron
interaction can induce some transitions of the ground state. It may be
possible to observe these transitions in the lower energy optical
absorption. Now, we present the FIR spectroscopy to elucidate such
transitions. Here, the phenomenological broadening parameter $\Gamma $ is
assumed to be $0.5$ {\rm meV}.

In Fig. 3, we plot the FIR absorption for the circularly polarized lights as
a function of the magnetic field. Compared with the no interaction case,
where the right and left circularly polarized (labeled by $\sigma _{\pm }$)
resonance energies are roughly given by $\omega _{\pm }=$ $\frac 12\left( 
\sqrt{\omega _c^2+4\omega _0^2}\pm \omega _c\right) $in the low field (as
shown in Fig. 4, $\sigma _{+}$ and $\sigma _{-}$ are represented by the
square and filled circle symbols, respectively), we find some unusual
features in the presence of interaction. ($i$) As the magnetic field
increases, the lowest right circularly polarized $\sigma _{+}$ resonance
energy shows discontinuous drops around $3$, $7$ and $10$ {\rm T} (see Fig.
3a). In contrast, this behavior is only found at $B\approx 8$ {\rm T} in the
case without the Coulomb interaction (see Fig. 4). Naturally, the
discontinuous drops are originated from the spin oscillation of the ground
state as mentioned above. Note that they are less reflected in the $\sigma
_{-}$ case. ($ii$) The absorption peak around $20$ {\rm meV} splits into
three or more subpeaks, especially in the $\sigma _{-}$ polarization case.
The striking behavior can be understood from the mixing of the
center-of-mass and the relative-motion modes. Certainly, it has to keep in
mind that due to the mixing there are many irregularly spaced and
near-degenerated energy levels hybridized from the relative-motion mode in
each $\left( S,L\right) $ subspace. The splitting can be identified as a
transition from the ground state to those ''hybrid'' states, showing its
two-electron characteristic. ($iii$) For both circular polarizations, the
lowest resonance energy shows slight blue shift due to the Coulomb
interaction.

Motivated by the above prominent features, we suggest that the spin
transitions of the ground state and the other effects induced by the
electron-electron interaction can be observed with circularly polarized
light. However, as pointed out by A. Lorke, in theirs experiment, the single
particle states are quite accurate basis for the description of the measured
FIR resonance. It seems in contrast to ours expectation. We argue here that
the controversy comes from the resolution of the experiment. With low
resolution, one is difficult to distinguish the details of the FIR
spectroscopy discussed above, and only the profile is explored. As shown in
Figs. (3a) and (3b), we can indeed observe that the profile of the FIR
resonance can be described by the single particle picture. Therefore, we
expect that our predictions about the electron-electron interaction may be
confirmed by a high resolution experiment in the future.

\section{Summary}

In conclusion, we have investigated the energy levels and far-infrared
spectroscopy of a two-electron nanoring in a magnetic field. Because of the
electron-electron interaction as well as the interplay between the magnetic
field and ring-like confinement potential, the nanoring exhibits rich
electronic structures. An obvious feature induced by the interaction is the
intersection between the lower levels. It presents the spin oscillation of
the ground state on the magnetic field. i.e., $\left( 0,0\right) \rightarrow
\left( 1,-1\right) \rightarrow \left( 0,-2\right) \rightarrow \left(
1,-3\right) $ states and so on. This phenomenon is indeed qualitatively
similar to that seen in a quantum dot, suggesting its intrinsic nature of
zero-dimensional quantum structures.

The profile of the FIR spectroscopy is roughly captured by the single
particle picture as indicated by a recent experiment. \cite{lorke} However,
the ring-like confinement potential doesn't allow the application of the
generalized Kohn theorem. Thus the single particle picture is inadequate for
seeing any effect due to electron-electron interaction. We have outlined
here two prominent features of the FIR spectroscopy caused by
electron-electron interaction: splitting and dis-continuous drops of the
resonance energies. We suggest that those two features can be detectable by
using the circularly polarized far-infrared light with high experimental
resolution.

\begin{center}
{\bf ACKNOWLEDGMENT}
\end{center}

The financial support from NSF-China (Grant No.19974019) and China's ''973''
program is gratefully acknowledged.

\begin{center}
{\bf Figures Captions}
\end{center}

Fig. 1. The energy levels of two electrons in a nanoring are plotted as a
function of the magnetic field with $\hbar \omega _0=12$ {\rm meV} and $%
R_0=14$ {\rm nm} in the (a) absence and (b) presence of electron-electron
interaction. The spin singlet and triplet states are labelled by solid and
dashed lines, respectively. The quantum number of each state (total spin,
total angular momentum) are also indicated. Note that only the lowest energy
level of each $\left( S,L\right) $ subspace is selected.

Fig. 2. The net Coulomb energies are plotted as a function of the magnetic
field. The other parameters are the same as in Fig. 1.

Fig. 3. A logarithmic 3D plot of the far-infrared absorption coefficient as
a function of the magnetic field for (a) right and (b) left circularly
polarized light.

Fig. 4. Far-infrared absorption resonance energies in the absence of
electron-electron interaction as a function of the magnetic field. The
square and filled circle symbols are corresponding to right and left
circularly polarized lights, respectively.

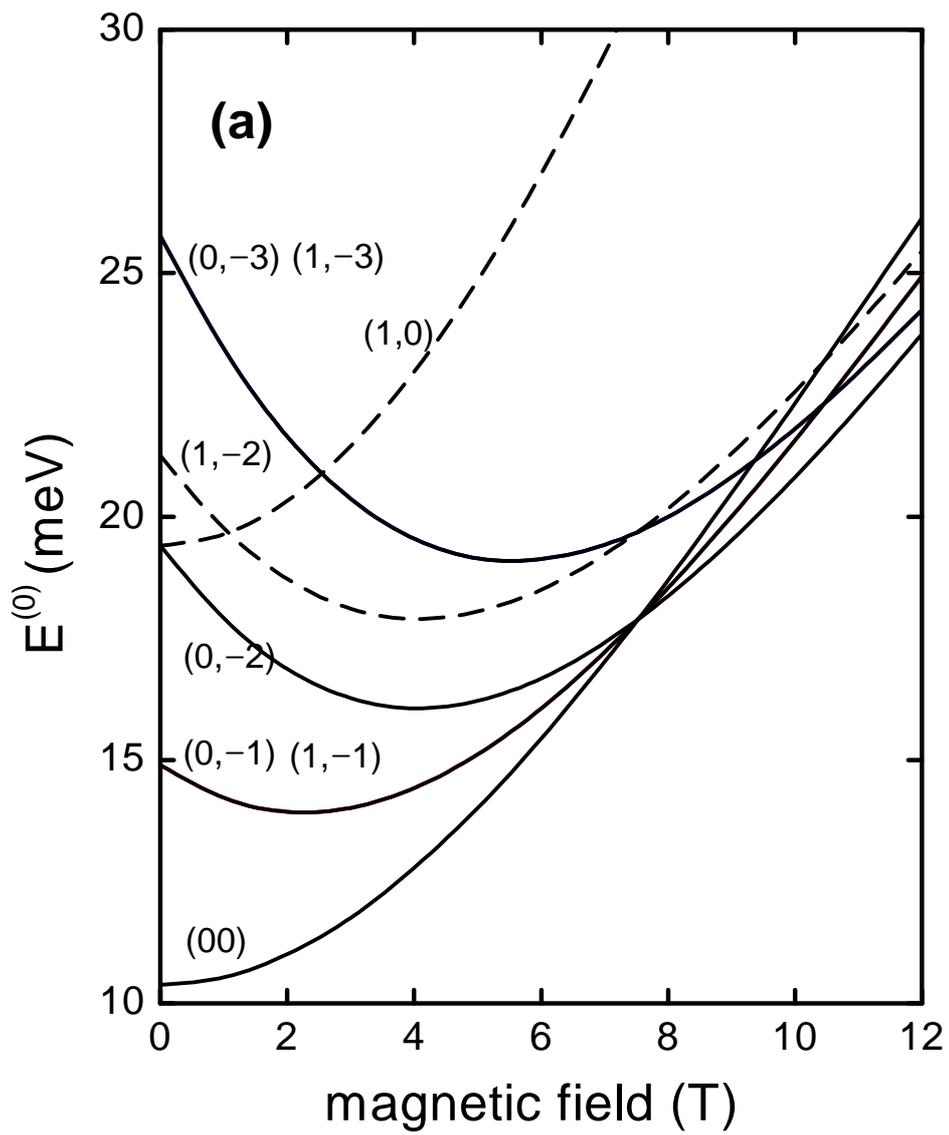
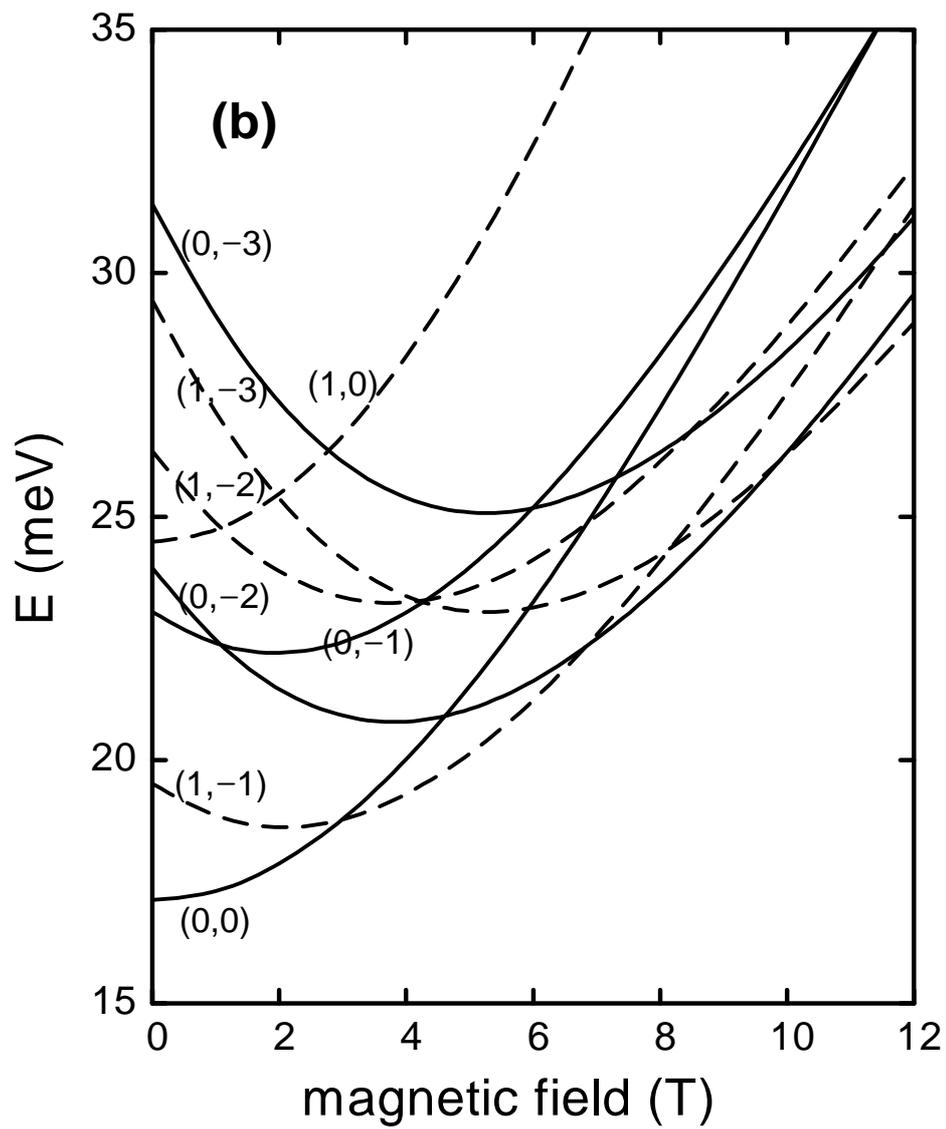
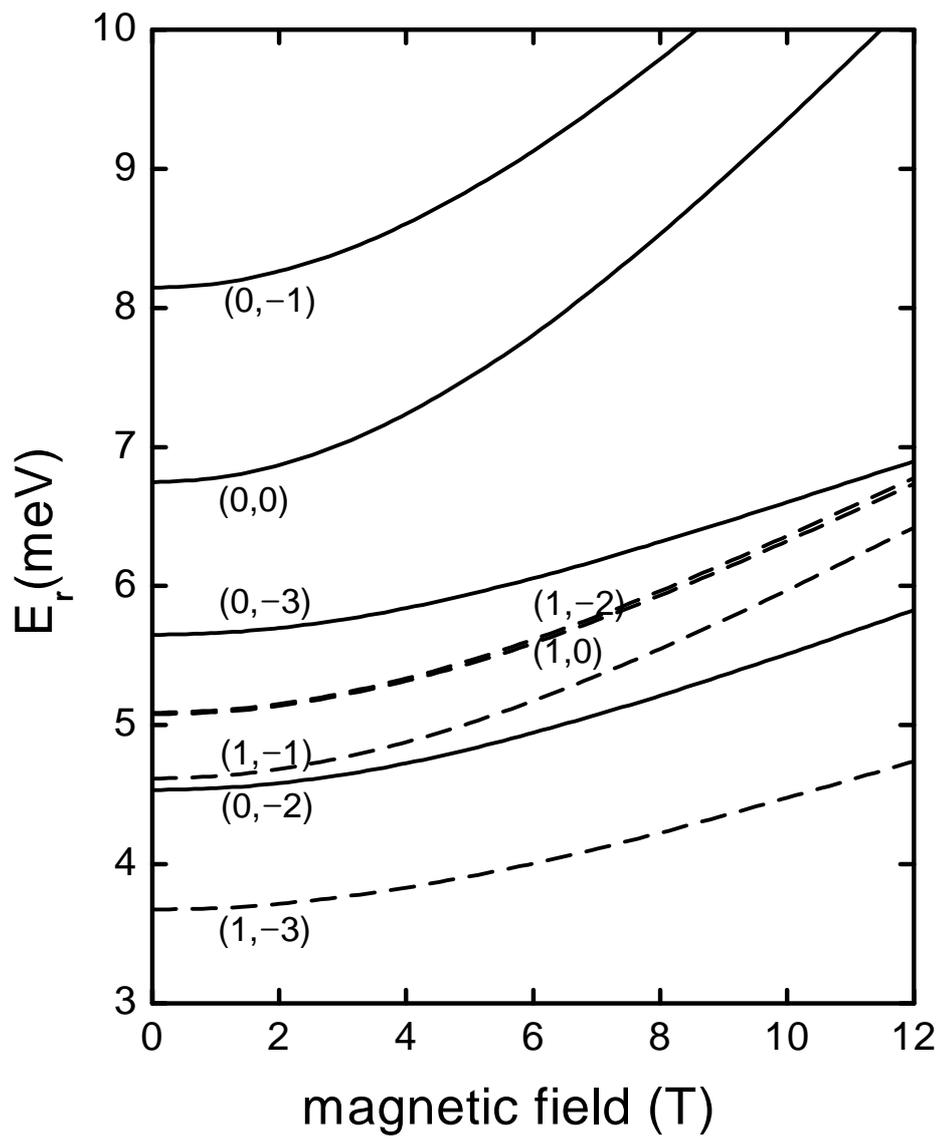
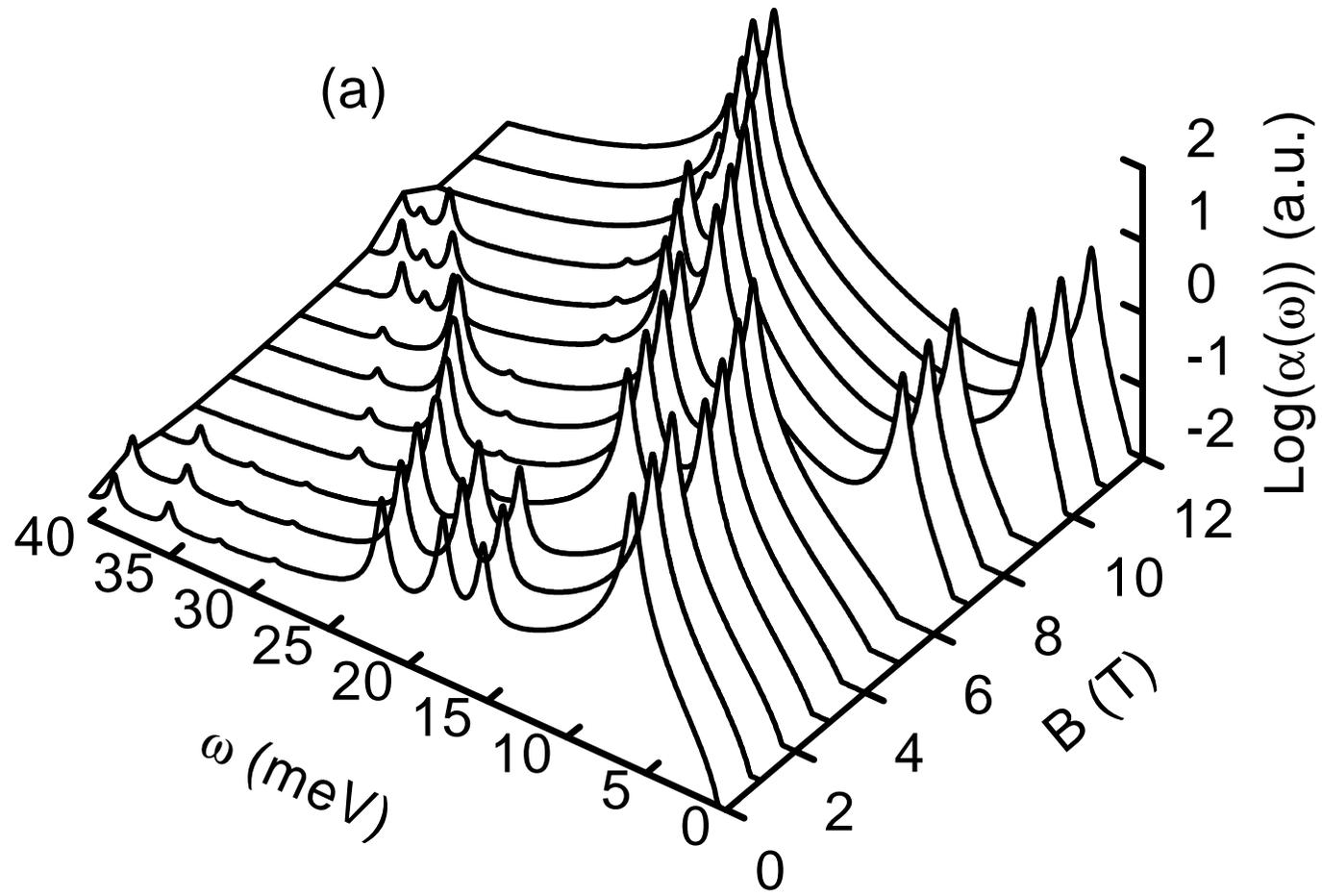
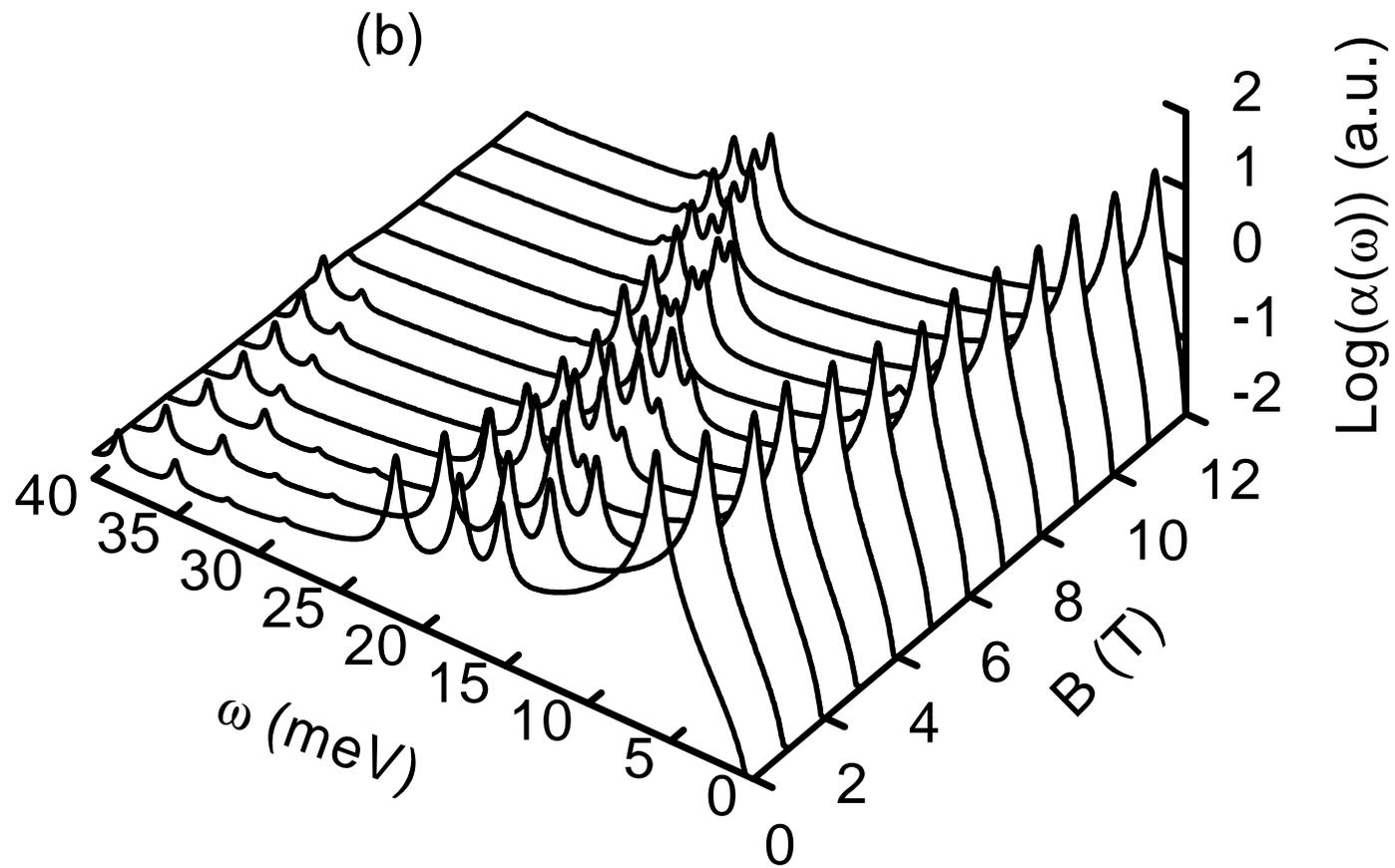
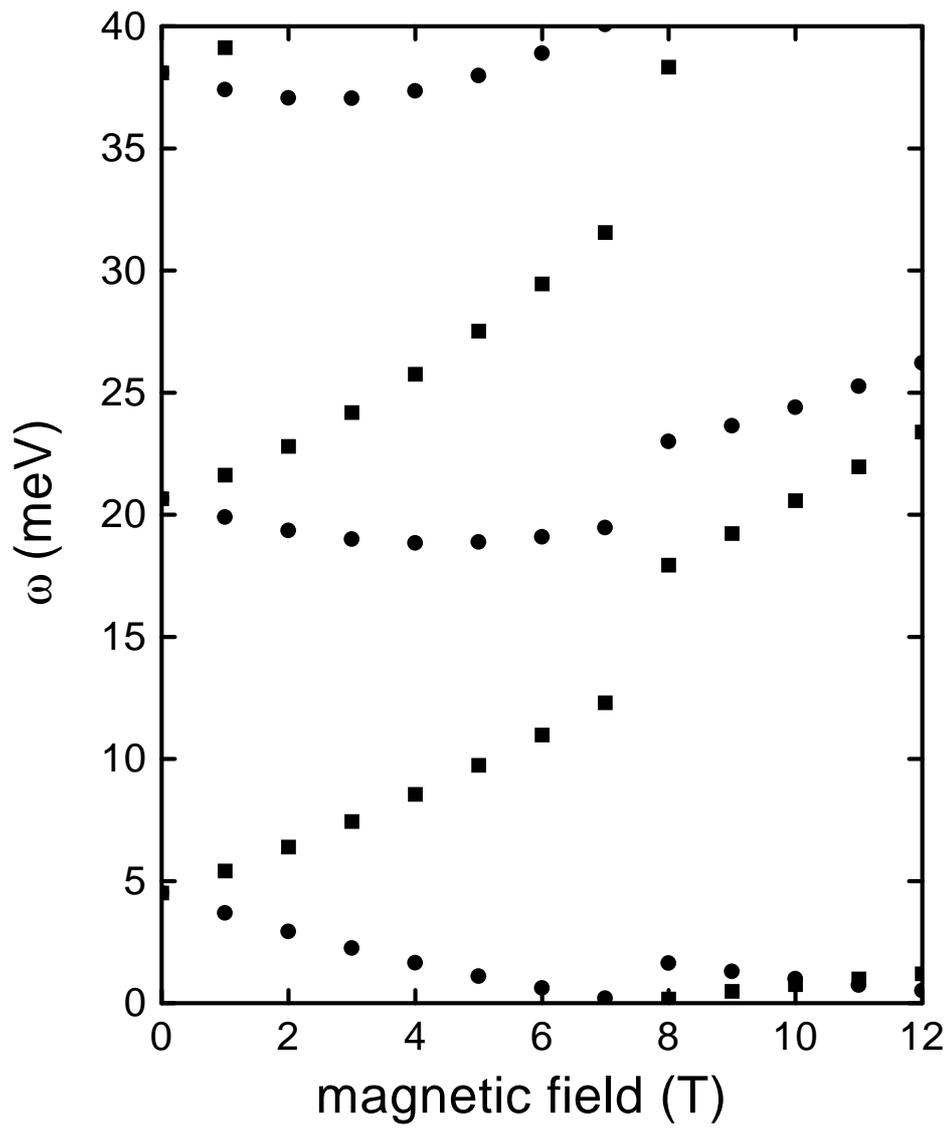
\end{document}